\documentclass{ws-ijmpd}
\usepackage[super,compress]{cite}
\begin{document}

\markboth{G. Cozzella, A. G. S. Landulfo, G. E. A. Matsas, D. A. T. Vanzella}
{QUEST FOR A ``DIRECT " OBSERVATION OF THE UNRUH EFFECT}

%
\catchline{}{}{}{}{}
%

\title{A QUEST FOR A ``DIRECT" OBSERVATION OF THE UNRUH 
EFFECT WITH CLASSICAL ELECTRODYNAMICS: IN HONOR OF ATSUSHI HIGUCHI
60$^{th}$ ANNIVERSARY}

\author{GABRIEL COZZELLA}

\address{Instituto de F\'\i sica Te\'orica, Universidade Estadual Paulista, R. Dr.\ Bento Teobaldo Ferraz, 271\\ 
S\~ao Paulo, S\~ao Paulo, 01140-070, Brazil\\
cozzella@ift.unesp.br}

\author{ANDR\'E G. S. LANDULFO}

\address{Centro de Ci\^encias Naturais e Humanas, Universidade Federal do ABC, Av. dos Estados, 5001\\
Santo Andr\'e, S\~ao Paulo,  09210-580, Brazil\\
andre.landulfo@ufabc.edu.br}

\author{GEORGE E. A. MATSAS}

\address{Instituto de F\'\i sica Te\'orica, Universidade Estadual Paulista, R. Dr.\ Bento Teobaldo Ferraz, 271\\ 
S\~ao Paulo, S\~ao Paulo, 01140-070, Brazil\\
matsas@ift.unesp.br}

\author{DANIEL A. T. VANZELLA}

\address{Instituto de F\'\i sica de S\~ao Carlos, Universidade de S\~ao Paulo, Caixa Postal 369\\ 
S\~ao Carlos, S\~ao Paulo, 13560-970, Brazil\\
vanzella@ifsc.usp.br}

\maketitle

\begin{history}
\received{Day Month Year}
\revised{Day Month Year}
\end{history}

\begin{abstract}
The Unruh effect is essential to keep the consistency of quantum field theory in inertial and uniformly accelerated 
frames.   Thus,  the Unruh effect must be considered as  well tested as  quantum field theory itself.  
In spite of it, it would be nice to realize an experiment whose output could be directly interpreted in terms of the Unruh 
effect.  This is not easy because the linear acceleration needed to reach a temperature 1 K is of order 
$10^{20}~{\rm m/s}^2$.  We discuss here a conceptually simple experiment reachable under present technology which may 
accomplish this goal. The inspiration for this proposal can be traced back to Atsushi Higuchi's Ph.D. thesis, 
which makes it particularly suitable to pay tribute to him on occasion of his $60^{\rm th}$ anniversary. 
\end{abstract}

\keywords{Unruh effect}

\ccode{PACS numbers: 04.62.+v}


\section{Introduction}	

In November 1972, Stephen Fulling submitted his famous paper~\cite{F73} 
``Nonuniqueness of canonical field quantization in Riemannian 
space-time", where he presented the Bogoliubov coefficients 
relating Minkowski and Rindler vacua and stated that {\em ``The 
notion of particle is completely different in the two theories"}. 
It would be shown later that this was closely connected to the 
seminal paper~\cite{H74} submitted by Stephen Hawking sixteen months later on the
evaporation of black holes. In the beginning, Hawking's 
effect was not consensual at all. The fact that Hawking's analysis 
implied an arbitrary large density of particles near the horizon 
was disturbing, since it seemed to impact on the hole's integrity 
itself. An intense discussion was on the verge of beginning. In August 
1974, Paul Davies submitted his paper~\cite{D75} ``Scalar particle production in 
Schwarzschild and Rindler metrics'', where the procedure used by 
Hawking in his 1975 paper~\cite{H75}, ``Particle creation by 
black holes'', was applied in flat spacetime to a Rindler 
coordinate system. In Ref.~\refcite{D75}, we find the famous acceleration 
temperature formula, 
\begin{equation}
    T_{\rm U}= a \hbar / (2 \pi k_{\rm B} c),
    \label{UnruhTemperature}
\end{equation}
but it was derived in the presence of a mirror and its physical content 
was quite unclear: ``The apparent production of particles is somewhat 
paradoxical, because there is no obvious source of energy for such a 
production". The fact that a uniformly accelerated observer 
with proper acceleration $a$ in the Minkowski vacuum sees a thermal 
bath of particles at a temperature~(\ref{UnruhTemperature}) 
was communicated by William Unruh~\footnote{Actually, this was clear 
to Unruh before April 1974 -- private communication.} in the $1^{\rm st}$ 
Marcel Grossmann meeting on General Relativity held in Trieste in July 1975. 
Unfortunately, the tradition of the Marcel Grossmann meeting of publishing 
proceedings late can be traced back to its first edition~\cite{U77}. 
In the meantime, Unruh published his renowned paper~\cite{U76} ``Notes on black hole 
evaporation" in 1976 announcing the effect named after him. 
His motivation was twofold. On the one hand, he wanted to understand better 
Fulling's 1972 result and on the other one Hawking's effect. It became clear, 
afterwards, that the thermal bath experienced by uniformly accelerated observers 
in Minkowski spacetime is composed of real particles (similarly to
the ones experienced by stationary observers outside evaporating black 
holes) as well as that their presence is consistent with a negligible 
backreaction effect. As a result, black holes would not be disrupted 
at all by the arbitrarily large temperature of Hawking radiation near its 
horizon. Interestingly enough, Geoffrey Sewell~\cite{S82} realized 
in 1982 that the Unruh effect was also codified in Bisognano 
and Wichmann 1976 work~\cite{BW76}. No matter how nonintuitive the Unruh 
effect is~\footnote{One could see ``accel. temp.'' listed among the four issues
 ``to learn'' at Richard Feynman's blackboard by the time he passed away in 1988 
 (see Fig.~1 of Ref.~\refcite{CHM08}).}, it should be clear since the mid 80s that it is 
necessary to keep the consistency of quantum field theory in inertial 
and uniformly accelerated frames and, thus, that it must be considered as well 
tested as quantum field theory itself~\cite{UW84} (see also 
Ref.~\refcite{VM01}). In spite of this, claims that the Unruh effect either 
does not exist or, more often, requires experimental confirmation 
can still be found in the literature (see, e.g., Ref.~\refcite{CM16} for 
a recent instance). Here, we bow in submission to what seems to be a 
demand to make the Unruh effect consensual, although some may see it as 
an unnecessary concession~\cite{PS14}.

Any observation of the Unruh effect must rely on resilient probes 
entrusted to (i)~record the Unruh effect and (ii)~make the 
information available to us, quasi-inertial observers.
The very origin of the difficulty of ``directly" 
observing the Unruh effect stems from the fact that the linear 
acceleration needed to reach a temperature $1~{\rm K}$ 
is of order~\cite{CHM08,FM14} $10^{20}~{\rm m/s}^2$.
Item~(i) drives our attention to massless rather than 
massive particles of the Unruh thermal bath. This is so because 
massive Rindler particles concentrate closer to the horizon than
massless ones, making their observation a harder 
enterprise~\cite{CCMV02,KSAD18}. This is why using accelerated 
observers to investigate, e.g., Planck scale particles (with 
masses of order $10^{19}~{\rm GeV}$), would be a terrible idea 
in practice. Among the massless particles, photons are much more 
promising than, say, gravitons, because their coupling to
physical detectors is typically much larger. Bell and Leinaas~\cite{BL83} 
were the first ones to say that the electron depolarization 
in storage rings could be explained in terms of the Unruh 
effect. They achieved partial success because 
strictly speaking the Unruh effect is not valid for circularly 
moving observers. Considering linear accelerators, rather than 
circular storage rings, is not an option, because the spins do 
not have enough time to thermalize in the Unruh thermal bath 
in the ultrashort lapse of time they get accelerated. In other
proposals~\cite{CT99,SSH06,OYZ16}, it is argued that one could relate the pairs 
of correlated photons emitted by accelerated charges and the 
corresponding charge quivering as seen by inertial observers 
with the scattering of Rindler photons of the Unruh thermal 
bath as defined by accelerated observers. 
The difficulty with  this strategy is that the radiation of such 
correlated photons, usually denominated quantum radiation, would 
require  still unavailable ultraintense lasers.

In a recent paper~\cite{CLMV17}, however, we have proposed a simple  
electromagnetic experiment feasible with present technology and
free of unfamiliar concepts, which should make clear that 
the Unruh thermal bath can be already seen in the much 
stronger signal of Larmor radiation (i.e., one-photon emission 
at the tree level), once one accepts the 
``indisputable" quantum formula: $E=\hbar \omega$. This is 
the single quantum ingredient we require to identify a signal 
of the Unruh effect in the classical Larmor radiation (see 
Ref.~\refcite{HM93} for a comprehensive discussion). The 
source of inspiration for such an experiment can be traced 
back to Ref.~\refcite{HMS92}, which, for its turn, is related 
to Atsushi Higuchi's Ph.D. paper~\cite{H87}, making it 
particularly suitable to pay tribute to him on occasion 
of his $60^{\rm th}$ anniversary. Here, we focus on the 
main challenges to realize such an experiment. 

The paper is organized as follows. In Sec.~\ref{The experimental 
apparatus} we present the experimental setup. In
Sec.~\ref{Experimental proposal to confirm the Unruh effect}
we explain the experimental strategy to confirm the Unruh effect.
Because, after all, the experiment is based on plain classical 
electrodynamics, we are confident to anticipate its output using 
Maxwell equations in Sec.~\ref{Virtual confirmation of the Unruh 
effect}. In Sec.~\ref{Experimental challenges} we discuss some 
experimental challenges to perform the real experiment. Our 
final conclusions are in Sec.~\ref{Conclusions}. We adopt metric 
signature $(+,-,-,-)$ and natural units, $G=c=k_{\rm B}=1$, unless 
stated otherwise.

\section{The experimental apparatus}
\label{The experimental apparatus}

Let us begin considering a pair of homogeneous and
constant electric and magnetic fields, $E^z $ and $B^z$, 
respectively, lying along the $z$ axis. We will assume here usual 
cylindrical coordinates $(t,z,r,\phi)$ as in Ref.~\refcite{CLMV17}. 
In general, a classical charge $q$ with mass $m_q$ in this environment will 
loop around the $z$ axis with some constant radius $r\equiv R$ and
non constant pitch because of the presence of the electric field. 
We assume that the charge moves initially against the direction
of the electric field, turning over at some point.
\begin{figure}[pb]
\centerline{\psfig{file=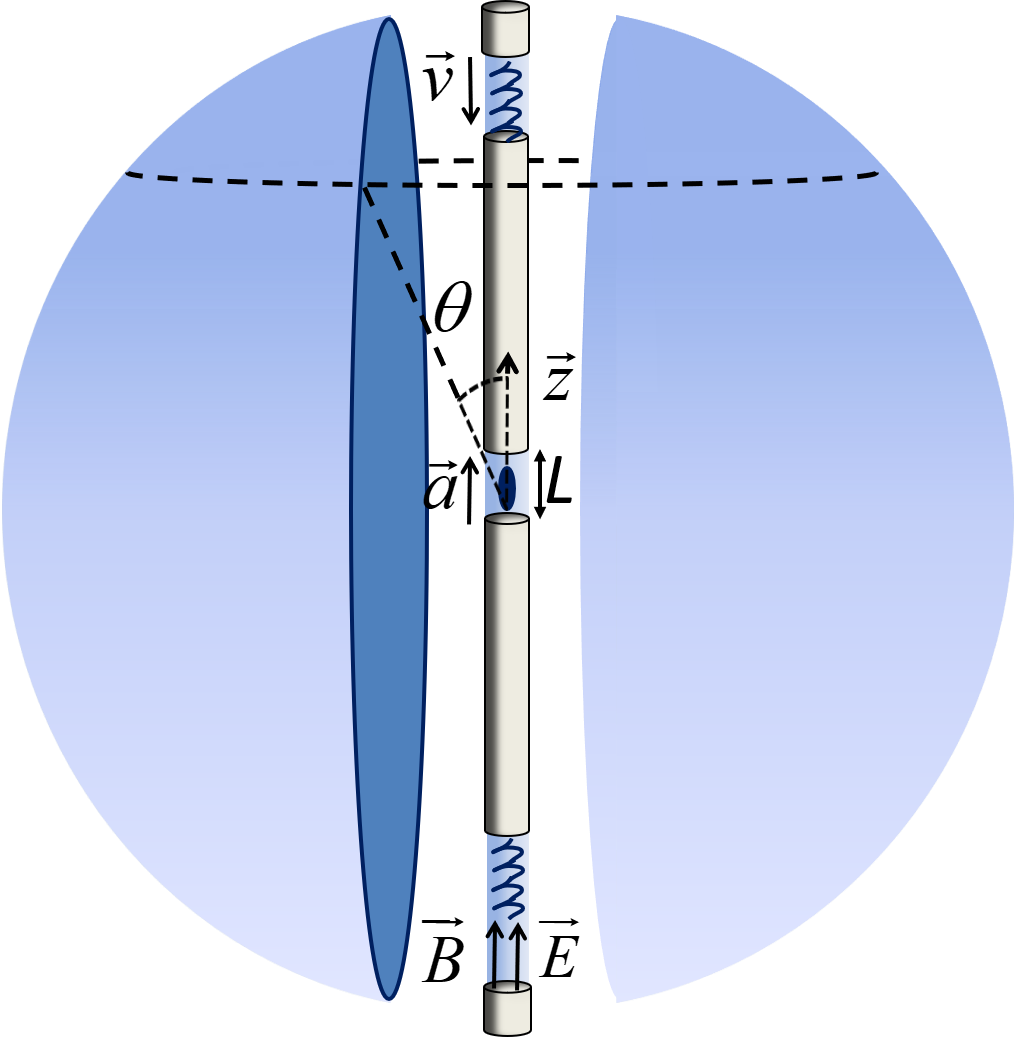, width=6.5cm}}
\vspace*{8pt}
\caption{Electrons are injected with velocity $\vec v$ in a cylinder 
containing linear electric and magnetic fields, $E^z$ and $B^z$, 
respectively. Radiation emitted where the charges make their U~turn 
(look at the center of the figure) is released through an open window and 
collected by detectors lying on the sphere. \label{Fig1}}
\end{figure}

A properly chosen linearly accelerated observer (named here, Atsushi, 
for brevity) moving along the $z$~axis according to the worldline
\begin{equation}
        t=a^{-1} \sinh{a\tau},\;\;\; 
        z=a^{-1} \cosh{a\tau},\;\;\;
        x=y=0,
\label{Rindler}
\end{equation}
with proper time $\tau$, and constant proper acceleration
$$
a = \frac{q E^z}{m_q \gamma},
$$
where $ \gamma \equiv 1/\sqrt{1-R^2 \Omega^2 }$,
will describe the charge as having closed circular trajectory with 
constant angular velocity
$$
\Omega \equiv \frac{d\phi}{d\tau} 
= \frac{B^z}{[m_q^2/q^2+ (R B^z)^2]^{1/2}}.
$$

A prototype experimental apparatus is shown in Fig.~\ref{Fig1}.  
The charges are injected in a cylinder containing the linear electric, 
$E^z$, and magnetic, $B^z$, fields. The radiation emitted 
where the charges make the U~turn is released  through an open window 
of length $L$ and collected by detectors set on a sphere with radius 
$R_S \gg L$. Now, since the Unruh effect concerns the Minkowski 
vacuum (free of boundaries), we must demand that the finite size of the 
window does not influence the results. Since the wavelengths of the 
emitted radiation goes as $\lambda \sim 1/a_{\rm tot}$, where 
$$
a_{\rm tot} = \gamma^2 \sqrt{a^2+ R^2 \Omega^4}
$$ 
is the charge {\it total} proper acceleration,
we must require 
\begin{equation}
a_{\rm tot} \sim 1/\lambda \gg 1/L 
\approx 10^{17}~{\rm m/s^2} \times (1~{\rm m}/L).    
\label{a_tot}
\end{equation}
\begin{figure}[pb]
\centerline{\psfig{file=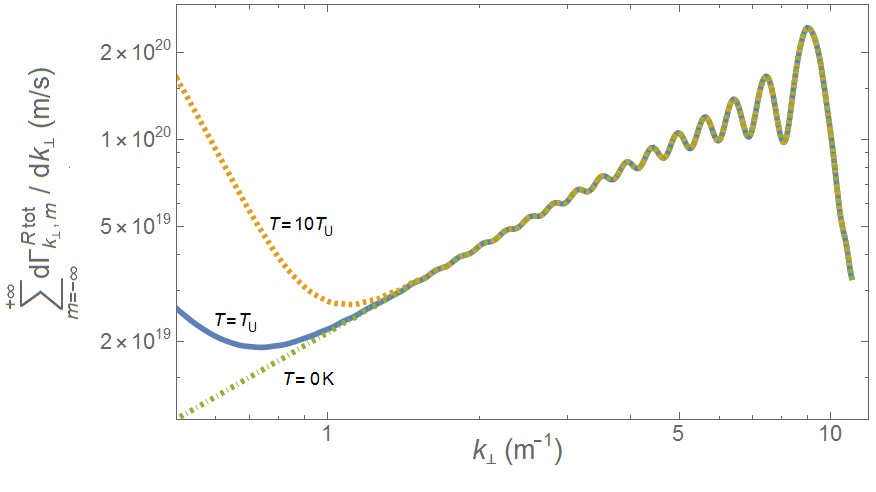, width=12cm}}
\vspace*{8pt}
\caption{ We plot ${dN_{k_\bot}^{{\rm R} }}/{dk_\bot}$  for different
    values of $T$ assuming $E^z = 1$~MV/m, $B^z = 10^{-1}$~T, $R = 10^{-1}$~m, 
   and injection energy $3.5$~MeV. The right-hand side of Eq.~(\ref{crucial}), 
   corresponding to the very prediction using the Unruh effect, is given by 
   the solid line.\label{Fig2}}
\end{figure}

Typical values achievable under present technology for the magnetic 
and electric fields~\cite{W08},  $B^z \approx  10^{-1}~{\rm T}$ 
and $E^z \approx 1~{\rm MV/m}$, respectively, produce accelerations 
$a \sim 10^{17}~{\rm m/s^2}$ and $a_{\rm tot}\sim 10^{19}~{\rm m/s^2}$,
where we have assumed $R\sim 10^{-1}~{\rm m}$. 

It is also worthwhile to note that we can neglect 
any radiation reaction effects on the charge trajectory. 
By using Larmor formula, we see that, under the conditions 
above, a bunch with a total charge of $10^7 e$ will emit (coherently) about
$
10 \times L/(1~{\rm m}) ~{\rm GeV},
$ 
which is, indeed, neglectable even in comparison to its 
rotational energy of about 
$
10^4 ~{\rm GeV}.
$

\section{Experimental proposal to confirm the Unruh effect} 
\label{Experimental proposal to confirm the Unruh effect} 

According to the Unruh effect, uniformly accelerated observers 
in the Minkowski vacuum will experience a thermal bath at the 
Unruh temperature. Then, Atsushi, who sees the charge making 
closed circles around him, shall also witness it emitting and 
absorbing Rindler photons to and from the underlying thermal 
bath, respectively. A straightforward calculation using quantum 
field theory in uniformly accelerated frames allows Atsushi 
to calculate the corresponding emission and absorption rates.
The combined distribution rate, i.e., emission plus absorption 
rates, of Rindler photons per  transverse momentum 
$k_\bot \in [0,+\infty)$ and fixed magnetic quantum number 
$m \in \mathbb{Z}$, is computed to be
\begin{eqnarray}
\frac{d\Gamma_{k_\bot m}^{{\rm R}\, {\rm tot}}}{dk_\bot}
&=& 
\frac {q^2 k_\bot}{ \pi^2 \hbar a}
\left[ \left| K'_{{im\Omega}/{a}} 
\left(\frac{k_\bot}{a}\right) \right|^2 | J_m (k_\bot R ) |^2 \right.
\nonumber \\
&+& 
\left.
(R \Omega)^2 
\left| 
K_{{i m \Omega}/{a}} 
\left(\frac{k_\bot}{a}\right) 
\right|^2
| J'_m (k_\bot R) |^2  
\right] \! 
\nonumber \\
& \times &
\sinh \left( \frac{\pi m \Omega}{a}\right)
\coth \left( \frac{ m \Omega \hbar}{2T}\right) \Theta(m),
\label{gammaRtotal}
\end{eqnarray}
where $J_n (x)$ and $K_n (x)$ are the first kind and 
modified Bessel functions, respectively, $``\, ' \,  "$ 
means derivative with respect to the argument, 
$\Theta (m) \equiv 0, 1/2$, and $1$ for $m<0, m=0$, 
and $m>0$, respectively, and $T$ is the temperature of 
the thermal bath. 

The issue to be experimentally decided is whether or 
not $T=T_{\rm U}$. To settle the issue, Atsushi uses the fact that 
{\em each absorption and emission of a Rindler photon 
from and to the Unruh thermal bath, respectively, 
will correspond to the emission of a Minkowski photon 
according to inertial observers}~\cite{UW84}. Therefore, 
the validity of the Unruh effect demands that inertial 
experimentalists must measure the following corresponding 
rate of emitted photons:
\begin{eqnarray}
& &  
\frac{dN_{k_\bot}^{{\rm M} }}{dk_\bot} 
\propto 
\left. 
\sum_{m=-\infty}^{\infty}
\frac{d\Gamma_{k_\bot m}^{{\rm R}\, {\rm tot} }}{dk_\bot} 
\right|_{T=T_{\rm U}}.
\label{crucial}
\end{eqnarray}
The proportionality  appears because the total number of 
emitted photons depends on how long the experiment is run. 
In Fig.~\ref{Fig2}, we plot 
$\sum_{m=-\infty}^{\infty}
{d\Gamma_{k_\bot m}^{{\rm R}\, {\rm tot} }}/{dk_\bot}$ 
for different values of $T$ assuming $E^z = 1$~MV/m, 
$B^z = 10^{-1}$~T, $R = 10^{-1}$~m, and injection energy 
$3.5$~MeV. The prediction given by Eq.~(\ref{crucial}) 
($T=T_{\rm U}$) is represented by the solid line. We recall 
that we should focus on the region $k_{\bot}\gtrsim 1/L$ 
in order to guarantee that the finite size of the window may 
be neglected. 
   
Rather than asking experimentalists to measure photons individually, 
it is enough to ask them to measure the spectral-angular distribution
$$
I(\omega, \theta,\phi)\equiv 
\frac{d{\cal E}(\omega, \theta,\phi)}{d\omega \,d(\cos\theta) \,d\phi}
$$ 
with ${\cal E}$ being the energy radiated away, since 
\begin{equation}
     \frac{dN_{k_\bot}^{{\rm M} } }{dk_\bot} 
     = \frac{k_\bot}{\hbar} \int_0^{2\pi} d\phi \int_{-\infty}^{\infty}  
     \frac{dk_z}{(k_\bot^2 + k_z^2)^{3/2}} I(\omega, \theta,\phi),
    \label{distribuicao inercial}
    \end{equation}
where
$\omega^2 = k_{\bot}^2+k_z^2$
and
$k_{\bot}=\omega \sin\theta$.
We note that the classical (Larmor) radiation corresponds to a coherent 
emission of photons and should be associated with a tree-level Feynman diagram 
with one single photon at the final state. We also note that the appearance of 
the $\hbar$ is thanks to the use of the 
one-photon relation ${\cal E}=\hbar \omega$. This is the crux of how the 
classical quantity $I(\omega, \theta,\phi)$ is translated into the 
quantum one $dN_{k_\bot}^{\rm M}/{dk_\bot}$ used to test the Unruh effect.

In summary, the idea is to perform the experiment above, 
measure $I(\omega, \theta,\phi)$, plug it in Eq.~(\ref{distribuicao inercial}),
and see whether or not $dN_{k_\bot}^{\rm M}/{dk_\bot}$ is in agreement 
with Eq.~(\ref{crucial}).

\section{Virtual confirmation of the Unruh effect}
\label{Virtual confirmation of the Unruh effect}

It happens, however, that we can anticipate instrumentalists 
and calculate the spectral-angular 
distribution
\begin{equation}
I(\omega, \theta,\phi) 
= \frac{R_S^2}{\pi} \left|\int_{-\infty}^{\infty} 
d t \ \vec{E}_{rad}(t,\theta,\phi) e^{-i \omega t} \right|^2 
\label{SpecDenERad}
\end{equation}
using Maxwell equations~\cite{Z12}, where 
$\vec{E}_{\rm rad}(t,\theta,\phi)$ is the electric field, 
which reaches the detectors on the sphere with radius $R_S$.

A straightforward calculation assuming our accelerated 
point-like charge and Eq.~(\ref{distribuicao inercial}) leads to 
\begin{equation}
\frac{dN_{k_\bot}^{{\rm M} }}{dk_\bot} 
 =
\left( \frac{4 \pi}{a} \int_{-\infty}^{\infty} 
\frac{d\kappa_z}{(1+\kappa_z^2)^{1/2}} \right)
\left. \sum_{m=-\infty}^{\infty}\frac{d\Gamma_{k_\bot  m}^{\rm R \, {\rm tot}}}{dk_\bot}\right|_{T=T_{\rm U}},
\label{final}
\end{equation}
which is in perfect agreement with Eq.~(\ref{crucial}). The fact 
that the term between parentheses diverges is because the 
calculation above has assumed a charge accelerating for infinite time, in 
which case an infinite number of photons is emitted for fixed $k_\bot$ 
element. Of course, in real experiments no divergence appears. 
\begin{figure}[pb]
\centerline{\psfig{file=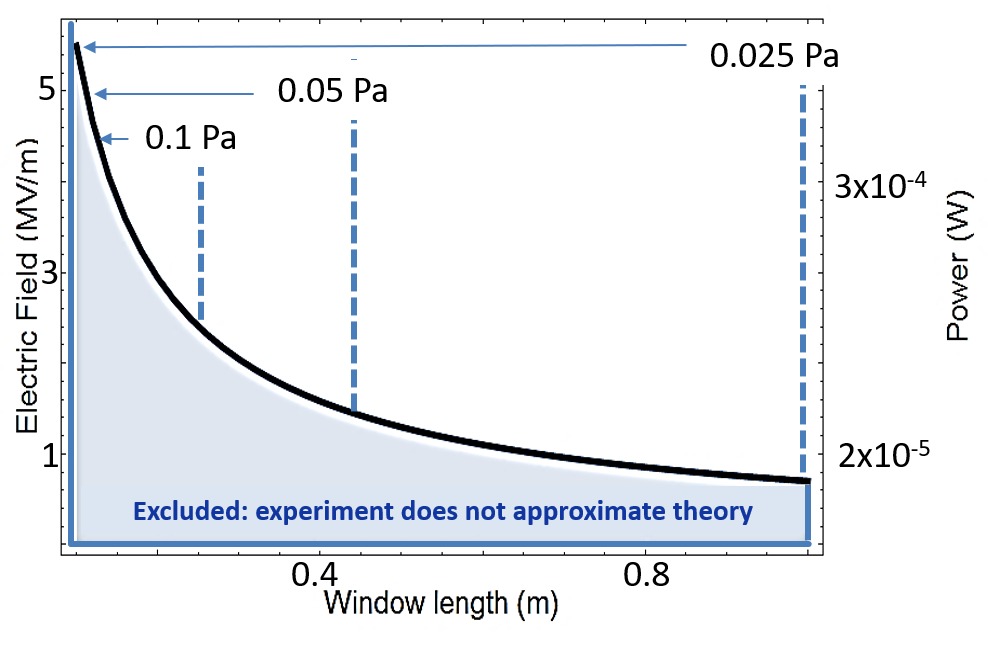, width=12cm}}
\vspace*{8pt}
\caption{ It is shown what is the minimum electric field 
    required in order that the typical photon wavelength be short
    enough to pass the window undisturbed by its finite size. It 
    also shows what is the vacuum level necessary in order to 
    avoid sparking and scattered of the charges by air molecules.  \label{Fig3}}
\end{figure}

\section{Do we need to perform the real experiment at all?} 
\label{Experimental challenges}

The natural question which follows is: {\em ``To what 
extent the theoretical calculation above can be seen as 
a virtual observation of the Unruh effect?"} A reasonable answer 
would be: {\em ``Since there is neither a reason to doubt Maxwell 
equations in the regime considered above nor any argument to distrust 
the golden quantum formula ${\cal E}=\hbar \omega$, the derivation 
above provides a sound anticipation of the output of the real 
physical experiment and, thus, it should be seen as a very  
confirmation of the Unruh effect}. 

On the other hand, we cannot dismiss the existence of ``hard-core" 
practitioners who may argue, e.g., that the fact that Maxwell 
equations have wonderfully succeeded 
up to now does not logically imply that, for some reason, they 
will not fail in the particular case here considered (no matter
how conservative is the regime where they are being applied). 
This would be as strange as raising the possibility that pink 
apples could float rather than fall down in the south pole 
against all odds just because such an experiment was never 
performed. Anyway, for these ones who argue 
in this way, we must concede that, to the best of our knowledge, 
no experiment has ever produced a graph like our Fig.~\ref{Fig2} 
for the Larmor radiation emitted by linearly accelerated charges 
with constant proper acceleration. This may sound surprising at 
first sight but it has an explanation: although doable 
under present technology, the proposed experiment is not trivial, 
as we will see next.
    
Before we begin discussing some experimental challenges to perform 
the experiment above, we introduce a simplification, namely, we 
set the magnetic field to zero. This drives us straightforwardly to 
Ref.~\refcite{HMS92}. In this instance, all radiation emitted as 
described by inertial observers corresponds to the emission and 
absorption of zero-energy Rindler photons as described by 
uniformly accelerated observers. In Ref.~\refcite{HMS92} a 
regularization procedure was implemented to deal with indeterminacies 
which appeared by the presence of these zero-energy Rindler photons. 
This was circumvented in Ref.~\refcite{CLMV17} by introducing the magnetic 
field, avoiding, thus, raising unnecessary concerns with
regularization procedures. Notwithstanding, because the crux of the
Unruh effect is the linear acceleration, this simplification
does not jeopardize the goal of this section by any means.

Firstly, we must recall that in order to avoid the 
finite size of the window to significantly influence 
the emitted radiation, we must keep a compromise 
between the window size and the minimum value of the 
electric field. Fig.~\ref{Fig3} plots the minimum electric 
field which is necessary to comply with Eq.~(\ref{a_tot}) 
as a function of the window size. E.g., for a 1~m window, 
the electric field should be at least about 1~MV/m.
\begin{figure}[pb]
\centerline{\psfig{file=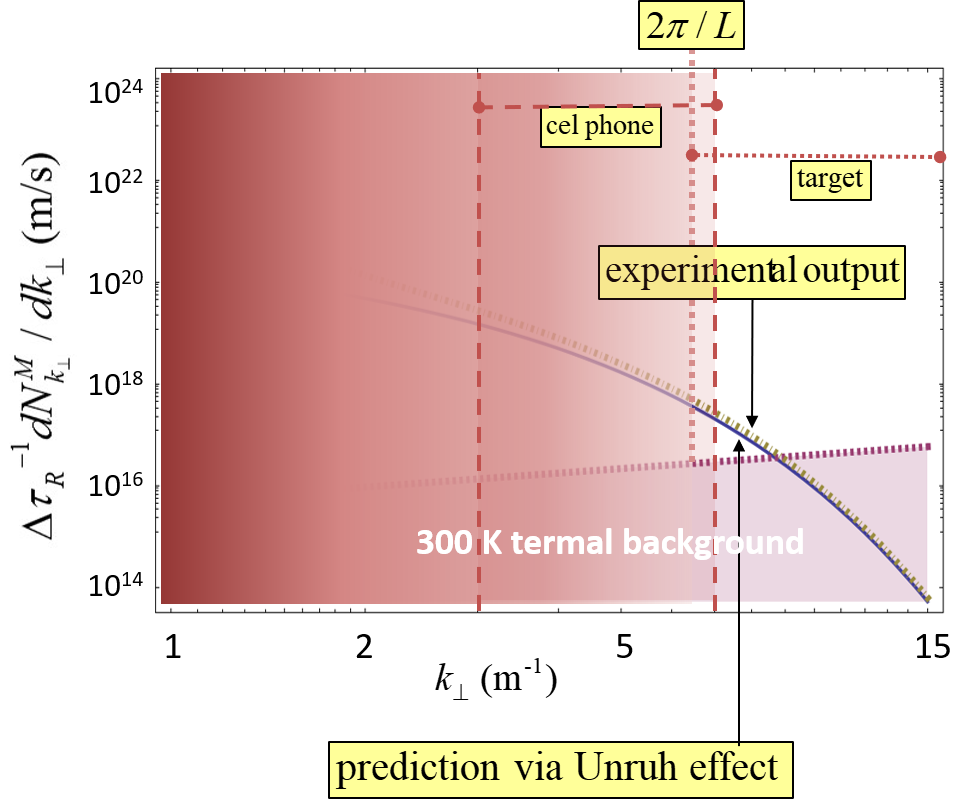, width=12cm}}
\vspace*{8pt}
\caption{It is shown the expected experimental output against the
 result obtained with the Unruh effect for a window size of $L= 1 {\rm m}$,
 an electric field $|E^z|= 1 {\rm MV/m}$,  a charged bunch containing
 $10^7~{\rm e}^-$, and detectors lying $R_S=10~{\rm m}$ far from the window.  \label{Fig4}}
\end{figure}

Next, we should guarantee the free mean path of the
charges to be at least larger than the window size. We do not
want our charges to be scattered by air molecules. In 
Fig.~\ref{Fig3} we also show the necessary vacuum to accomplish 
this. It is clear 
that making vacuum is not an issue. A trivial vacuum of 
0.025~Pa would be enough to perform the experiment with 
windows as large as 1~m long.

From Fig.~\ref{Fig3} we may be induced to believe that mild electric 
fields would be just fine, once the window is chosen large enough. 
This is not the case, however, because the detectors must lie at the 
radiation zone, where, strictly speaking, $R_S \gg L$. In 
Fig.~\ref{Fig4}, we show the expected photon detection rate per 
transverse momentum $k_\bot$ assuming a window size $L= 1 {\rm m}$, 
an electric field $|E^z|= 1 {\rm MV/m}$, and a charged bunch containing 
$10^7~{\rm e}^-$, which should be measured by detectors lying 
$R_S=10~{\rm m}$ far from the window. (The plot in Fig.~\ref{Fig4} 
differs from the one shown in Fig.~\ref{Fig2} because of the absence 
of the magnetic field.) In Fig.~\ref{Fig4}, the left-hand side of the 
vertical dotted line $k_\bot=2 \pi /L$ was excluded because the 
experiment would be sensitive to the window size. We also exclude
signal contamination due to a background $300~{\rm K}$ thermal bath
and cell-phone communication (assuming Brazilian regulation). At the 
end, we are left with a small observation region, where, nevertheless, 
we expect 
(i)~the prediction performed via the Unruh effect and 
(ii)~the experimental signal to fit each other very well.

\section{Conclusions}  
\label{Conclusions} 

We have discussed a doable experiment whose output can be directly interpreted in 
terms of the Unruh effect. Although it is based 
on simple classical electrodynamics, it has not been performed yet. Rather than 
waiting experimentalists to do it, we have calculated the output theoretically and
obtained full agreement with the Unruh effect. This should be enough for most 
scientists to consider it as a virtual observation of the Unruh effect; {\em 
who does doubt Maxwell equations when applied to regular regimes?} Despite it, 
we cannot dismiss the existence of radical contenders arguing, e.g., that the 
fact that Maxwell equations have worked so nicely until today does not imply, 
strictly speaking, they will work as desired in the particular case considered
here {\em no matter how conservative is the regime where they are being applied}. 
For these ones, we dedicate Sec.~\ref{Experimental challenges}, where
the challenges to realize the physical experiment above are discussed.

\section*{Acknowledgments}

We are deeply indebted to Steve Fulling, Atsushi Higuchi, Bill Unruh, and 
Bob Wald for sharing some of their recollections. We are also thankful to
the referee for comments. G.M. would also like to thank Daniel Sudarsky for 
various discussions on the Unruh effect along the years. G.~C. and  
A.~L., G.~M., D.~V. acknowledge  S\~ao Paulo Research Foundation (FAPESP) 
for full and partial support  under Grants 2016/08025-0 and 2017/15084-6, 
2015/22482-2, 2013/12165-4, respectively. G.~M. was also partially supported 
by Conselho Nacional de Desenvolvimento Cient\'\i fico e Tecnol\'ogico (CNPq).


\end{document}